\documentclass[twocolumn,showpacs,preprintnumbers,amsmath,amssymb]{revtex4}
\usepackage{graphicx}
\usepackage{textcomp}

\begin{document}
\draft
\title{Bloch oscillations in tight-binding lattices with defects}
  \normalsize

\author{Stefano Longhi\footnote{Author's email address:
longhi@fisi.polimi.it}}
\address{Dipartimento di Fisica, Politecnico di Milano, Piazza L. da Vinci 32, I-20133 Milano, Italy}


%
\bigskip
\begin{abstract}
\noindent Defects in tight-binding lattices generally destroy the
onset of Bloch oscillations (BOs) and the periodic self-imaging of
the wave packet due to the lack of an equally-spaced Wannier-Stark
ladder spectrum. Here it is shown that localized and extended
defects in the lattice can be engineered to be transparent for BOs.
Such lattices are synthesized from the defect-free lattice by the
technique of intertwining operators generally employed in
supersymmetric quantum mechanics. The energy spectrum of the
synthesized lattices differs from the Wannier-Stark ladder of the
defect-free lattice because of the missing of pairs of resonances in
the ladder, thus ensuring the persistence of BOs.
\end{abstract}


\maketitle

\section{Introduction}
Bloch oscillations (BOs) represent a fundamental coherent transport
phenomenon originally predicted for quantum particles in periodic
potentials driven by an external dc force. BOs have been observed in
a wide variety of quantum and classical physical systems, including
semiconductor superlattices \cite{Waschke93}, ultracold atoms
\cite{ultracold}, Bose-Einstein condensates \cite{BEC}, arrays of
evanescently-coupled optical waveguides
\cite{OBOwaveguides1,OBOwaveguides2}, photonic superlattices
\cite{OBOsuperlattices},  and acoustical superlattices
\cite{acoustical}. In the absence of dephasing and scattering
processes and for a negligible Zener tunneling, the particle
accelerated by the external force undergoes an oscillatory (rather
than a translational) motion owing to Bragg scattering off the
lattice. The periodicity of the motion is basically ascribed to the
transition of the energy spectrum from a continuous band with
delocalized Bloch eigenstates in absence of the external force, to a
discrete ladder spectrum with localized Wannier-Stark (WS)
eigenstates when the external force is applied and Zener tunneling
is negligible. A single-band tight-binding lattice in one spatial
dimension provides the simplest model to describe the onset of BOs
and the formation of WS localized states \cite{Hartmann04,note0}.
The inclusion of lattice disorder, nonlinearities, inhomogeneities
or defects has generally a detrimental effect on BOs. Several
authors have investigated the onset of BOs in generalized
tight-binging lattice models. Among others, we mention BOs in
nonlinear lattices \cite{Bishop96} and in aperiodic lattices with
long-range correlated disorder \cite{Adame}, BOs with
spatially-inhomogeneous dc fields \cite{Sacchetti08}, BOs in
quasicrystals \cite{Farinato00}, BOs for interacting bosons
\cite{Kolovsky03}, BOs in disordered lattices with interparticle
interaction \cite{Lewenstein08}, and BOs in lattices with
inhomogeneous intersite couplings \cite{Longhi09}. The introduction
of localized or extend defects in the lattice, for either the
intersite couplings or the site energies, is generally expected to
destroy the WS ladder spectrum of the defect-free lattice, and thus
the onset of BOs. In this work it is shown that localized and
extended defects in the lattice can be engineered to be transparent
for BOs. Such lattices are synthesized starting from the defect-free
lattice Hamiltonian by application of the technique of intertwining
operators, successfully employed in supersymmetric quantum mechanics
\cite{susy1,susy2} to add or delete energy states of a given
Hamiltonian. In our case, the spectrum of the synthesized lattices
differ from the Wannier-Stark ladder of the defect-free lattice
because of the missing of pairs of resonances in the ladder, which
ensures the persistence of BOs and self-imaging of a wave packet.
The paper is organized as follows. In Sec.II, the technique of
intertwining operators for spectral engineering of tight-binding
Hamiltonians is presented. In Sec.III, the technique is then applied
to synthesize lattice models with defects that support BOs. Finally,
Sec.IV outlines the main conclusions.\\

\section{Tight-binding lattice engineering}
\par Let us consider a one-dimensional tight-binding lattice described by the Hamiltonian
\begin{equation}
\mathcal{H}= \sum_{n}  \kappa_{n} \left( |n-1\rangle \langle n|+|n
\rangle \langle n-1|\right)  + \sum_n V_n |n \rangle \langle n|
\end{equation}
where $|n\rangle$ is a Wannier state localized at site $n$ of the
lattice, $\kappa_n$ is the hopping rate between sites $|n-1 \rangle$
and $|n \rangle$, and $V_n$ is the energy of Wannier state $|n
\rangle$ in presence of the external applied force. To ensure the
Hermiticity of $\mathcal{H}$, the hopping amplitudes $\kappa_n$ and
site energies $V_n$ must assume
 real values. For a defect-free lattice and a homogenous force, the
intersite coupling $\kappa_n$ is independent of $n$
($\kappa_n=\kappa$) and the site energy $V_n$ increases linearly
with $n$, i.e. $V_n=-Fn$, where $F$ is the potential gradient. The
corresponding tight-binding Hamiltonian will be denoted in the
following by $\mathcal{H}_0$ and referred to as the WS Hamiltonian.
As is well known, its spectrum is purely discrete and the allowed
energy levels are given by $E_l=l F$, where $l=0, \pm 1, \pm 2, ...$
\cite{note0}. Our goal is to synthesize a lattice Hamiltonian, of
the form given by Eq.(1), with an energy spectrum which differs from
that of the WS Hamiltonian $\mathcal{H}_0$ because of the missing of
some resonances in the WS ladder spectrum. This spectral engineering
problem can be solved by the technique of intertwining operators
generally employed in supersymmetric quantum mechanics \cite{susy1},
which is also valid for matrix Hamiltonians (see, for instance,
\cite{susy2} and references therein). The technique is briefly
described in a rather
general manner in this section.\\
Let us indicate by $\mathcal{H}_1$ the tight-binding Hamiltonian
defined by Eq.(1) with hopping amplitudes and site energies given by
$\{ \kappa^{(1)}_n, V^{(1)}_n \}$, and let us assume that $\kappa_n
\rightarrow \kappa$ as $n \rightarrow \pm \infty$ and that $|V_n|$
is bounded or diverges with an algebraic law $|V_n| \sim n$  as $ n
\rightarrow \pm \infty$; note that such assumptions are satisfied
for the WS Hamiltonian $\mathcal{H}_0$. Let us indicate by $\mu_1$,
$\mu_2$, $\mu_3$, ... the point spectrum of $\mathcal{H}_0$ and let
$|\phi^{(1)} \rangle=\sum_n \phi^{(1)}_n |n \rangle $ be the
(proper) eigenfunction of $\mathcal{H}_1$ corresponding to the
energy $\mu_1$, i.e. \cite{note1}
\begin{equation} \kappa^{(1)}_{n}
\phi_{n-1}^{(1)}+\kappa^{(1)}_{n+1} \phi_{n+1}^{(1)}+V_{n}^{(1)}
\phi_{n}^{(1)} =\mu_1 \phi_{n}^{(1)},
\end{equation}
 with $|\phi_{n}^{(1)}| \rightarrow 0$ for $n \rightarrow  \pm \infty$. It can
be readily shown that, provided that $\phi_n^{(1)}$ does not vanish,
the following factorization for $\mathcal{H}_1$ holds
\begin{equation}
\mathcal{H}_1=\mathcal{Q}_1 \mathcal{R}_1+\mu_1
\end{equation}
where
\begin{eqnarray}
\mathcal{Q}_1 & = & \sum_n \left( q_{n}^{(1)} |n \rangle \langle
n|+\bar{q}_{n-1}^{(1)} |n-1 \rangle \langle n| \right) \\
\mathcal{R}_1 & = & \sum_n \left( r_{n}^{(1)} |n \rangle \langle
n|+\bar{r}_{n+1}^{(1)} |n+1 \rangle \langle n| \right)
\end{eqnarray}
and
\begin{eqnarray}
r_n^{(1)} & = & - \sqrt{\frac{\kappa_n^{(1)}
\phi_{n-1}^{(1)}}{\phi_n^{(1)}}} \\
\bar{r}_n^{(1)} & = & -\frac{\kappa_{n}^{(1)}}{r_{n}^{(1)}} \\
q_n^{(1)} & = & -r_n^{(1)} \\
\bar{q}_n^{(1)} & = & - \bar{r}_{n+1}^{(1)}
\end{eqnarray}
Let us then construct the new Hamiltonian $\mathcal{H}_2$ obtained
from $\mathcal{H}_1$ by interchanging the operators $\mathcal{R}_1$
and $\mathcal{Q}_1$, i.e. let us set
\begin{equation}
\mathcal{H}_2=\mathcal{R}_1 \mathcal{Q}_1+\mu_1.
\end{equation}
Using  Eqs.(4-9), from E.(10) it can be readily shown that
$\mathcal{H}_2$ describes the Hamiltonian of a tight-binding lattice
[i.e., it is of the form (1)] with hopping amplitudes and site
energies $\{ \kappa^{(2)}_{n},V^{(2)}_n\}$ given by
\begin{eqnarray}
\kappa^{(2)}_n & = & \kappa_n^{(1)} \frac{r^{(1)}_{n-1}}{r^{(1)}_n} \\
V_n^{(2)} & = & V_n^{(1)}+\kappa_{n+1}^{(1)}
\frac{\phi_{n+1}^{(1)}}{\phi_n^{(1)}}-\kappa_n^{(1)}
\frac{\phi^{(1)}_n}{\phi^{(1)}_{n-1}}.
\end{eqnarray}
An interesting property of the new Hamiltonian $\mathcal{H}_2$ is
that its energy spectrum is the same as that of $\mathcal{H}_1$,
apart from the lack of the discrete energy level $E=\mu_1$. In fact,
let us indicate by $| \psi_E \rangle=\sum_n \psi_n(E) | n \rangle$ a
proper (or improper) eigenfunction of $\mathcal{H}_1$ with energy
$E$. Note that, if $E$ belongs to the point spectrum of
$\mathcal{H}_1$, $|\psi_n(E)| \rightarrow 0$ as $n \rightarrow \pm
\infty$, whereas if  $E$ belongs to the continuous spectrum of
$\mathcal{H}_1$, $|\psi_n(E)|$ remains bounded as $n \rightarrow \pm
\infty$. Let us first assume that $E \neq \mu_1$. Using the
factorization (3) for $\mathcal{H}_1$, the eigenvalue equation
$\mathcal{H}_1 | \psi_E \rangle = E | \psi_E \rangle$ reads
explicitly
\begin{equation}
\mathcal{Q}_1 \mathcal{R}_1 | \psi_E \rangle=(E-\mu_1) | \psi_E
\rangle
\end{equation}
from which it follows that $\mathcal{R}_1| \psi_E \rangle \neq 0$
since $E \neq \mu_1 $. Applying the operator $\mathcal{R}_1$ to both
sides of Eq.(13), one obtains
\begin{equation}
\mathcal{R}_1 \mathcal{Q}_1 | \tilde{\psi}_E \rangle=(E-\mu_1) |
\tilde{\psi}_E \rangle,
\end{equation}
i.e. $\mathcal{H}_2 | \tilde{\psi}_E \rangle= E | \tilde{\psi}_E
\rangle$, where we have set $ | \tilde{\psi}_E \rangle=\mathcal{R}_1
|\psi_E \rangle$ or, explicitly [see Eq.(5)]
\begin{equation}
\tilde{\psi}_n(E)=r_n^{(1)} \psi_n(E)+\bar{r}_n^{(1)} \psi_{n-1}(E).
\end{equation}
Therefore, $ | \tilde{\psi}_E \rangle$ is an eigenfunction of
$\mathcal{H}_2$ corresponding to the energy $E$. Also, from Eqs.(6),
(7), (15) and from the assumed asymptotic behavior of
$\kappa_n^{(1)}$ and $V_n^{(1)}$, it follows that $|\tilde{\psi}_E
\rangle$ is a proper (improper) eigenfunction of $\mathcal{H}_2$ in
the same way as $|\psi_E \rangle$ is a proper (improper)
eigenfunction of $\mathcal{H}_1$. In a similar way, one can show
that any eigenvalue $E$ of $\mathcal{H}_2$, belonging to the
continuous or to the point spectrum (with $E \neq \mu_1$), is also
an eigenvalue of $\mathcal{H}_2$. Therefore the continuous and point
spectra of $\mathcal{H}_1$ and  $\mathcal{H}_2$ do coincide, apart
from the $E=\mu_1$ eigenvalue which needs a separate analysis. For
$E=\mu_1$, the (proper) eigenfunction of $\mathcal{H}_1$ is by
construction $| \phi^{(1)} \rangle$ [see Eq.(2)], which satisfies
the condition $\mathcal{R}_1 | \phi^{(1)} \rangle=0$. On the other
hand, from Eq.(10) it follows that the eigenvalue equation
$\mathcal{H}_2 |\psi \rangle = \mu_1 | \psi \rangle$ is satisfied
for either $|\psi \rangle = | \psi^{(1)} \rangle$ or $|\psi \rangle
= | \psi^{(2)} \rangle$, where $\mathcal{Q}_1 | \psi^{(1)} \rangle
=0$ and $\mathcal{Q}_1 | \psi^{(2)} \rangle =| \phi^{(1)} \rangle$.
 The equation $\mathcal{Q}_1 | \psi^{(1)} \rangle
=0$ reads explicitly
\begin{equation}
q_n^{(1)} \psi_n^{(1)}+{q}_n^{(1)} \psi_{n+1}^{(1)}=0.
\end{equation}
Using the expressions of $q_n^{(1)}$ and $\bar{q}_n^{(1)}$ given by
Eqs.(6-9), the difference equation (16) for $\psi_n^{(1)}$ can be
solved in a closed form, yielding
\begin{equation}
\psi_n^{(1)}=\frac{1}{\sqrt{\kappa_n^{(1)} \phi^{(1)}_n
\phi^{(1)}_{n-1} }}.
\end{equation}
In view of the asymptotic behavior of $\phi_n^{(1)}$ and $\kappa_n$
as $n \rightarrow \pm \infty$, it turns out that $\psi_n^{(1)}$ is
unbounded as $ n \rightarrow \pm \infty$, i.e. it is not an
eigenfunction (neither proper not improper) of $\mathcal{H}_2$.
Similarly, as $\phi^{(1)}_n \rightarrow 0$ as $n \rightarrow \pm
\infty$, the equation $\mathcal{Q}_1 | \psi^{(2)} \rangle =|
\phi^{(1)} \rangle$ for $| \psi^{(2)} \rangle$ reduces to Eq.(16) in
the asymptotic limit $ n \rightarrow \pm \infty$, and thus also
$\psi_n^{(2)}$ is unbounded at $ n \rightarrow \pm \infty$.
Therefore, none of the two linearly independent solutions $|
\psi^{(1)} \rangle$ and $| \psi^{(2)} \rangle$ of the second-order
difference equation $\mathcal{H}_2 | \psi \rangle = \mu_1 |\psi
\rangle$ are bounded, i.e. $\mu_1$
does not belong neither to the point spectrum nor to the continuous spectrum of $\mathcal{H}_2$.\\
The factorization method can be iterated to construct new
Hamiltonians $\mathcal{H}_3$, $\mathcal{H}_4$, $\mathcal{H}_5$, ...
whose energy spectra differ from that of $\mathcal{H}_1$ owing to
the missing of the discrete energy levels $\{ \mu_1, \mu_2 \}$, $\{
\mu_1, \mu_2, \mu_3 \}$, $\{ \mu_1, \mu_2, \mu_3, \mu_4  \}$, ...
For instance, to construct the Hamiltonian $\mathcal{H}_3$, let
$|\psi \rangle$ be the (proper) eigenfunction of $\mathcal{H}_1$
corresponding to the energy $E=\mu_2$, and let us set $| \phi^{(2)}
\rangle=\mathcal{R}_1 | \psi \rangle$. From the previous analysis,
it follows that $|\phi^{(2)} \rangle$ is the (proper) eigenfunction
of $\mathcal{H}_2$ corresponding to the energy $E=\mu_2$, i.e.
\begin{equation}
\kappa^{(2)}_{n} \phi_{n-1}^{(2)}+\kappa^{(2)}_{n+1}
\phi_{n+1}^{(2)}+V_{n}^{(2)} \phi_{n}^{(2)} =\mu_2 \phi_{n}^{(2)}.
\end{equation}
Let us then construct the new operators $\mathcal{R}_2$ and
$\mathcal{Q}_2$, defined as in Eqs.(4-9) but with $\phi^{(1)}_n$ and
$\kappa_n^{(1)}$ replaced by $\phi^{(2)}_n$ and $\kappa_n^{(2)}$,
respectively. The factorization $\mathcal{H}_2=\mathcal{Q}_2
\mathcal{R}_2+\mu_2$ then holds. Reversing the order of the
$\mathcal{R}$ and $\mathcal{Q}$ operators, one obtains the new
Hamiltonian $\mathcal{H}_3=\mathcal{R}_2 \mathcal{Q}_2+\mu_2$, which
possesses  the same energy spectrum of $\mathcal{H}_0$, except for
the missing of the two energy levels $E=\mu_1$ and $E=\mu_2$. The
hopping amplitudes $\kappa_n^{(3)}$ and site energies $V_n^{(3)}$ of
the lattice corresponding to the Hamiltonian $\mathcal{H}_3$ are
given by Eqs.(11) and (12), with $\phi_n^{(1)}$, $r_n^{(1)}$,
$\kappa_n^{(1)}$ and $V_n^{(1)}$ replaced by $\phi_n^{(2)}$,
$r_n^{(2)}$, $\kappa_n^{(2)}$ and $V_n^{(2)}$, respectively. It
should be noted that the technique of intertwining operators so far
described could generate lattice Hamiltonians with complex-valued
hopping rates $\kappa_n$ or site energies $V_n$; for instance, the
hopping amplitudes of $\mathcal{H}_2$ might become complex-valued
when $\phi_n^{(1)}$ does not have a defined sign [see Eq.(6) and
(11)]. This situation, corresponding to a non-Hermitian lattice with
real-valued energy spectrum, will not be considered in this work. It
should be nevertheless observed that iteration of the intertwining
operator technique could finally restore the Hermiticity of the
lattice Hamiltonian, in spite some of the intermediate Hamiltonians
are not self-adjoint. This is precisely the case of our interest
that will be discussed in the next section.

\section{Bloch oscillations}
Let us consider the WS Hamiltonian $\mathcal{H}_0$, defined by
Eq.(1) with $\kappa_n=\kappa$ and $V_n=-Fn$. As is well known (see,
e.g., \cite{Hartmann04}), $\mathcal{H}_0$ has a purely point
spectrum (the WS ladder spectrum) with energies [see Fig.1(a)]
\begin{equation}
E_l=l F
\end{equation}
 and corresponding localized eigenstates
\begin{equation}
 |u^{(l)} \rangle=\sum_n J_{n+l}(\gamma) |n \rangle,
\end{equation}
\begin{figure}[htbp]
  \includegraphics[width=85mm]{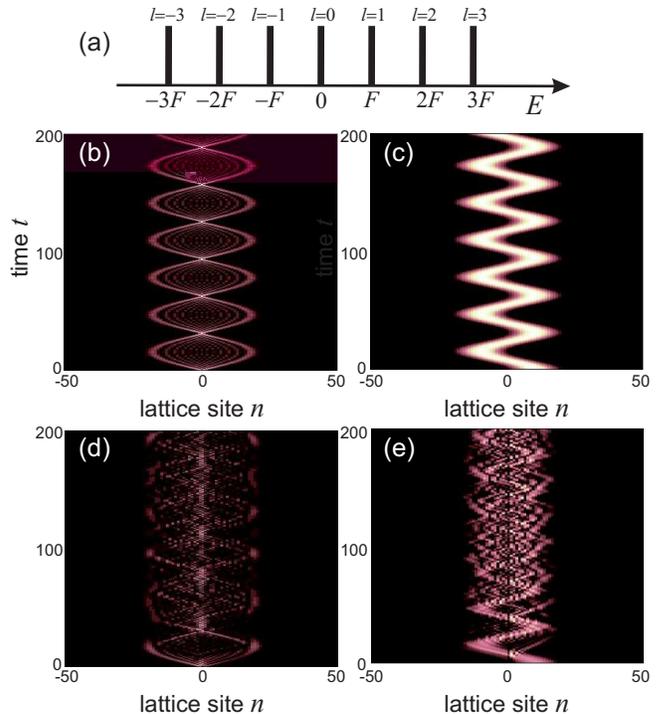}\\
   \caption{(color online) (a) Schematic of the energy levels (WS ladder) of a
   WS Hamiltonian $\mathcal{H}_0$ ($\kappa_n=\kappa$, $V_n=-Fn$); the energy level spacing is uniform and equal to $F$.
   (b) and (c) Periodic breathing and oscillatory modes in
   a WS Hamiltonian corresponding to $F=0.2$, $\kappa=1$ and to single-site
   excitation [$c_n(0)=\delta_{n,0}$ in (b)] and broad Gaussian wave packet excitation [$c_n(0)=\exp[-(n-10)^2/64]$ in (c)] of the lattice at initial time $t=0$.
   Note the periodic self-imaging of time
   evolution at multiplies of the BO period $T_B=2 \pi/ F$. (d) and (e): same as (b) and (c), but
   for the WS Hamiltonian with a defect in the hopping rate ($\kappa_n=1$
   for $n \neq 1$, $\kappa_1=1.5$).}
\end{figure}
\begin{figure}[htbp]
  \includegraphics[width=85mm]{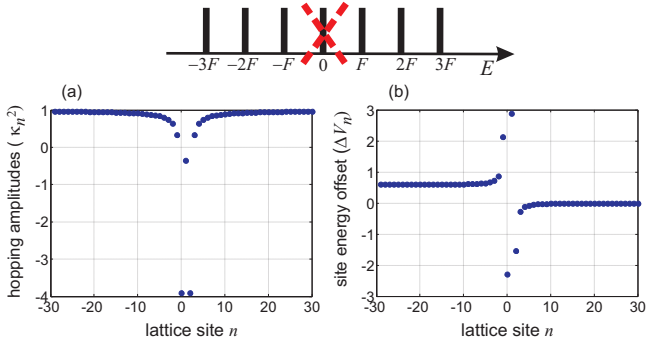}\\
   \caption{(color online) Behavior of (a) the square of hopping amplitudes $[\kappa_n^{(2)}]^2$, and
(b) of site energy offset $\Delta V_n^{(2)}=V_n^{(2)}+Fn$ for the
Hamiltonian $\mathcal{H}_2$ synthesized from the WS Hamiltonian
$\mathcal{H}_0$ by removal of the WS resonance $\mu_1=0$ (see the
inset). Parameter values are $\kappa=1$ and $F=0.6$.}
\end{figure}
  where $\gamma=2 \kappa /F$ and $l=0, \pm 1, \pm 2, ...$ is the quantum number \cite{note0}.
   Note that the WS state with quantum number $l$ is localized at around the lattice site
  $n=-l$. Owing to the equal spacing of WS modes, the temporal evolution of any wave
packet is periodic and self-imaging is attained at times multiples
of the Bloch period $T_B=2 \pi / F$. This effect is clearly visible
by observing the breathing or oscillatory modes \cite{Hartmann04}
corresponding to either an initial single-site or broad-site
excitation of the lattice, as shown in Figs.1(b) and (c). The
introduction of some defects in the lattice, in either the hopping
amplitudes $\kappa_n$ or site energies $V_n$, generally breaks the
self-imaging property of the lattice [see, as an example, the
simulations shown in Figs.1(d) and (e)]. In this section we aim to
construct tight-binding lattices with defects in which the
self-imaging phenomenon of the WS (defect-free) Hamiltonian
$\mathcal{H}_0$ is maintained. This goal can be achieve by the
application of the intertwining operator technique described in the
previous section assuming as the starting Hamiltonian
$\mathcal{H}_1$ the WS Hamiltonian $\mathcal{H}_0$, i.e.
$\mathcal{H}_1=\mathcal{H}_0$. The main idea is that any new
Hamiltonian $\mathcal{H}_2$, $\mathcal{H}_3$, $\mathcal{H}_4$, ...
obtained from $\mathcal{H}_0$ by successive application of the
intertwining operator technique has a spectrum which differs from
that of $\mathcal{H}_0$ by the missing of some of the WS resonances,
and thus a periodic temporal dynamics of the wave packet is
maintained with the same period $T_B$ of the original WS
Hamiltonian. As a first step, let us construct the Hamiltonian
$\mathcal{H}_2$ by removing from the spectrum of the WS Hamiltonian
$\mathcal{H}_0$ one WS resonance, for instance the one with energy
$\mu_1=0$ corresponding to the quantum number $l=0$ in Eq.(19).
According to Eqs.(11) and (12) with $k_n^{(1)}=\kappa$,
$V_n^{(1)}=-Fn$ and $\phi^{(1)}_n=J_n(\gamma)$, the hopping
amplitudes and site energies of the lattice associated to
$\mathcal{H}_2$ read explicitly
\begin{eqnarray}
\kappa_n^{(2)} & = & \kappa
\sqrt{\frac{J_{n}(\gamma)J_{n-2}(\gamma)}{J^2_{n-1}(\gamma)}}
\\
V_n^{(2)} & = & -Fn+\kappa
\frac{J_{n+1}(\gamma)}{{J_{n}(\gamma)}}-\kappa
\frac{J_{n}(\gamma)}{{J_{n-1}(\gamma)}}.
\end{eqnarray}
\begin{figure}[htbp]
  \includegraphics[width=85mm]{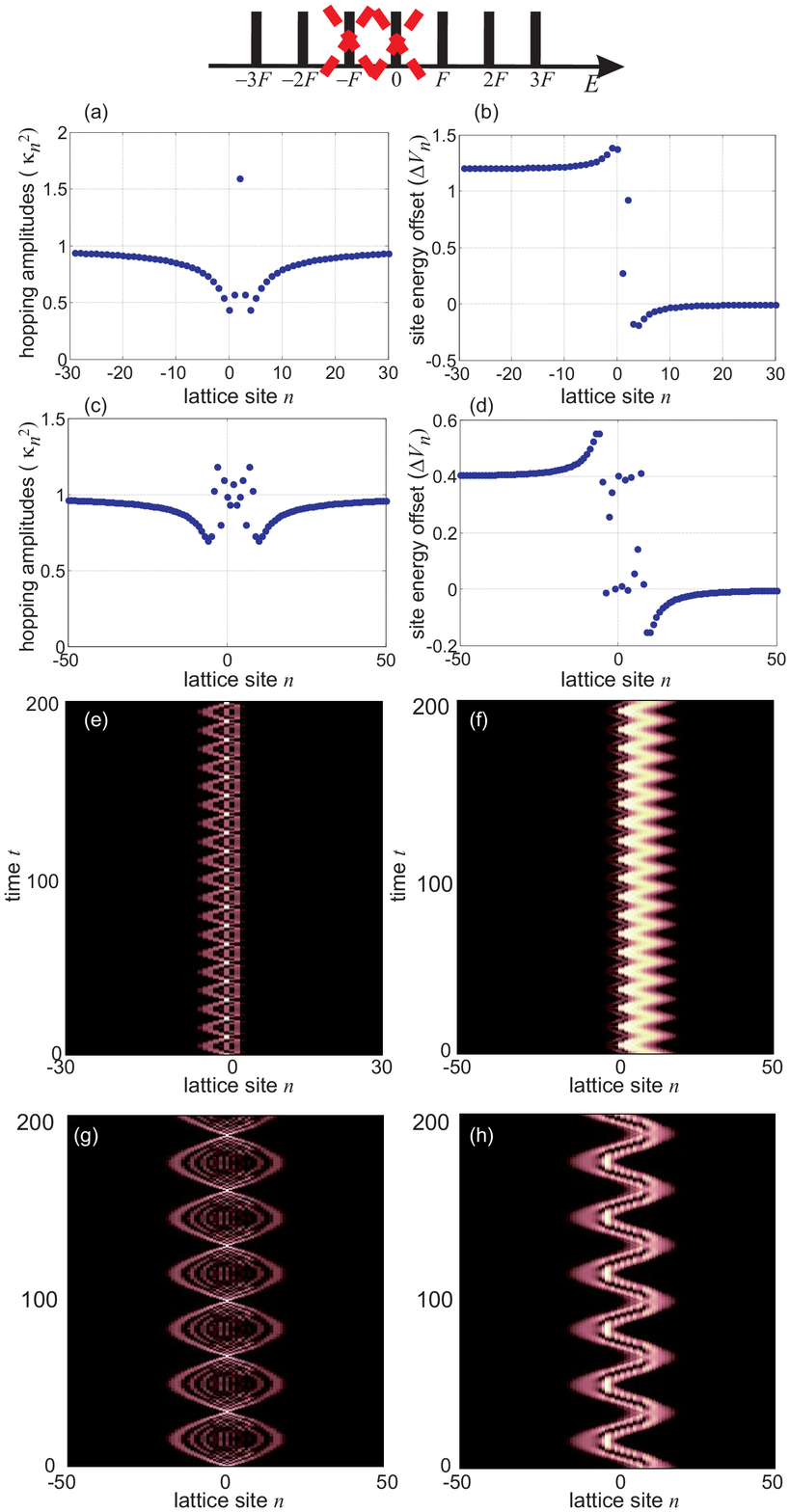}\\
   \caption{(color online) (a-d) Behavior of the square of hopping amplitudes $[\kappa_n^{(3)}]^2$ and
of site energy offset $\Delta V_n^{(3)}=V_n^{(3)}+Fn$ for the
Hamiltonian $\mathcal{H}_3$ synthesized from the WS Hamiltonian
$\mathcal{H}_0$ by removal of the WS resonances $\mu_1=0$ and
$\mu_2=-F$ (see the inset at the top of the figure). Parameter
values are $\kappa=1$ and $F=0.6$ in (a) and (b), and $\kappa=1$ and
$F=0.2$ in (c) and (d). (e) and (f): persistent BOs (breathing and
oscillatory modes) in the the synthesized lattice corresponding to
$F=0.6$. (g) and (h): Same as (e) and (f), but for the synthesized
lattice with $F=0.2$.}
\end{figure}

\begin{figure}[htbp]
  \includegraphics[width=85mm]{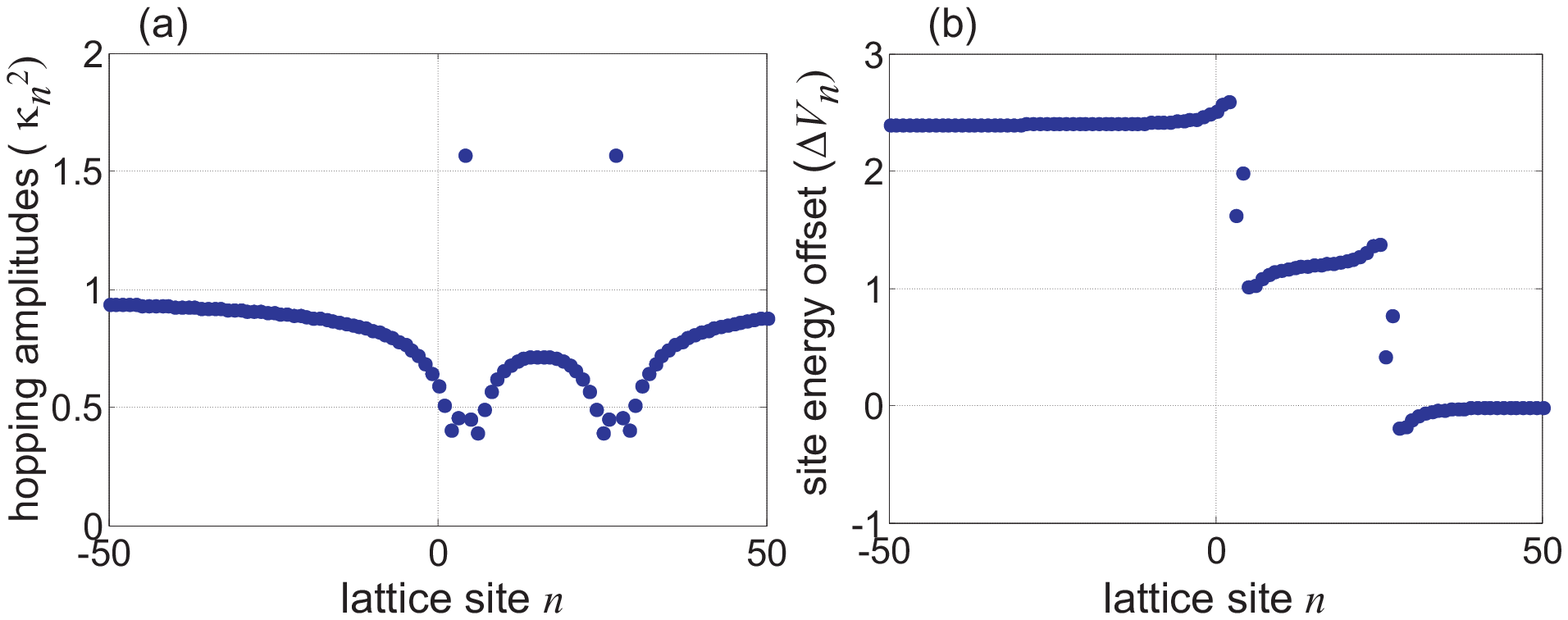}\\
   \caption{(color online) Same as Fig.2, but for the Hamiltonian $\mathcal{H}_5$ synthesized from $\mathcal{H}_0$ by
   removing the four WS resonances $E=0,-F,-25F,-26F$. Parameter values are $\kappa=1$ and $F=0.6$.}
\end{figure}
A typical behavior of $[\kappa_n^{(2)}]^2$ and $\Delta V_n^{(2)}
\equiv V_n^{(2)}-V_n^{(1)}=V_n^{(2)}+Fn$, as predicted by Eqs.(21)
and (22), is shown in Fig.2. Note that $\kappa^{(2)}_n \rightarrow
\kappa$ and $\Delta V^{(2)}_n$ settles down to constant values as $n
\rightarrow \pm \infty$, so that far from the defect near $n=0$ the
lattice described by $\mathcal{H}_2$ behaves like a defect-free WS
lattice. Note also that, owing to the asymptotic behavior of Bessel
functions $J_n$ at large indices, one has $\Delta V_n^{(2)}
\rightarrow 0$ as $n \rightarrow + \infty$ but $\Delta V_n^{(2)}
\rightarrow 2 \kappa /\gamma=F$ as $n \rightarrow -\infty$ [see
Fig.2(b)]. Unfortunately, due to the oscillating behavior of Bessel
functions $J_n(\gamma)$, the hopping rates $\kappa_n^{(2)}$ can
become complex-valued and the Hamiltonian $\mathcal{H}_2$,
correspondingly, ceases to be non-Hermitian. This is clearly shown
in Fig.2(a), where $[\kappa_n^{(2)}]^2$ becomes negative at a few
lattice sites near $n=0$. This circumstance indicates that, for such
indices, the hopping rates are purely imaginary. In spite of the
non-Hermiticity of $\mathcal{H}_2$, its spectrum remains real-valued
and BOs with the same period $T_B$ as that of the WS Hamiltonian are
found \cite{note2}. Fortunately, a second application of the
intertwining operator technique, assuming as a second energy level
$\mu_2$ of $\mathcal{H}_0$ one of the two WS resonances adjacent to
$\mu_1$, i.e. $\mu_2= \pm F$ [corresponding to the quantum number
$l= \pm 1$ in Eq.(19)], the Hermiticity of $\mathcal{H}_3$ is
restored. As an example, Fig.3 shows the behaviors of
$[\kappa_n^{(3)}]^2$ and $\Delta V_n^{(3)} \equiv
V_n^{(3)}-V_n^{(1)}$ corresponding to $\mu_2=-F$ and for two values
of $\gamma$. Note that, for $n \rightarrow \pm \infty$,
$\kappa_n^{(3)} \rightarrow 1$ and $\Delta V_n^{(3)}$ settles down
to constant values [$\Delta V_n^{(3)} \rightarrow 0$ for $n
\rightarrow +\infty$, $\Delta V_n^{(3)} \rightarrow 2F$ for $n
\rightarrow -\infty$], so that the synthesized lattice described by
$\mathcal{H}_3$ behaves like the WS (defect-free) lattice for site
indices far from $n=0,1$, i.e. far from the lattice sites of WS
localized states removed by the intertwining operator technique. As
$F$ is decreased, i.e. the localization length of the WS states is
increased [see (20)], the localization length of the defect in the
lattice is increased, as one can see by comparing Figs.3(a),(b) with
Figs.3(c),(d).
\begin{figure}[htbp]
  \includegraphics[width=85mm]{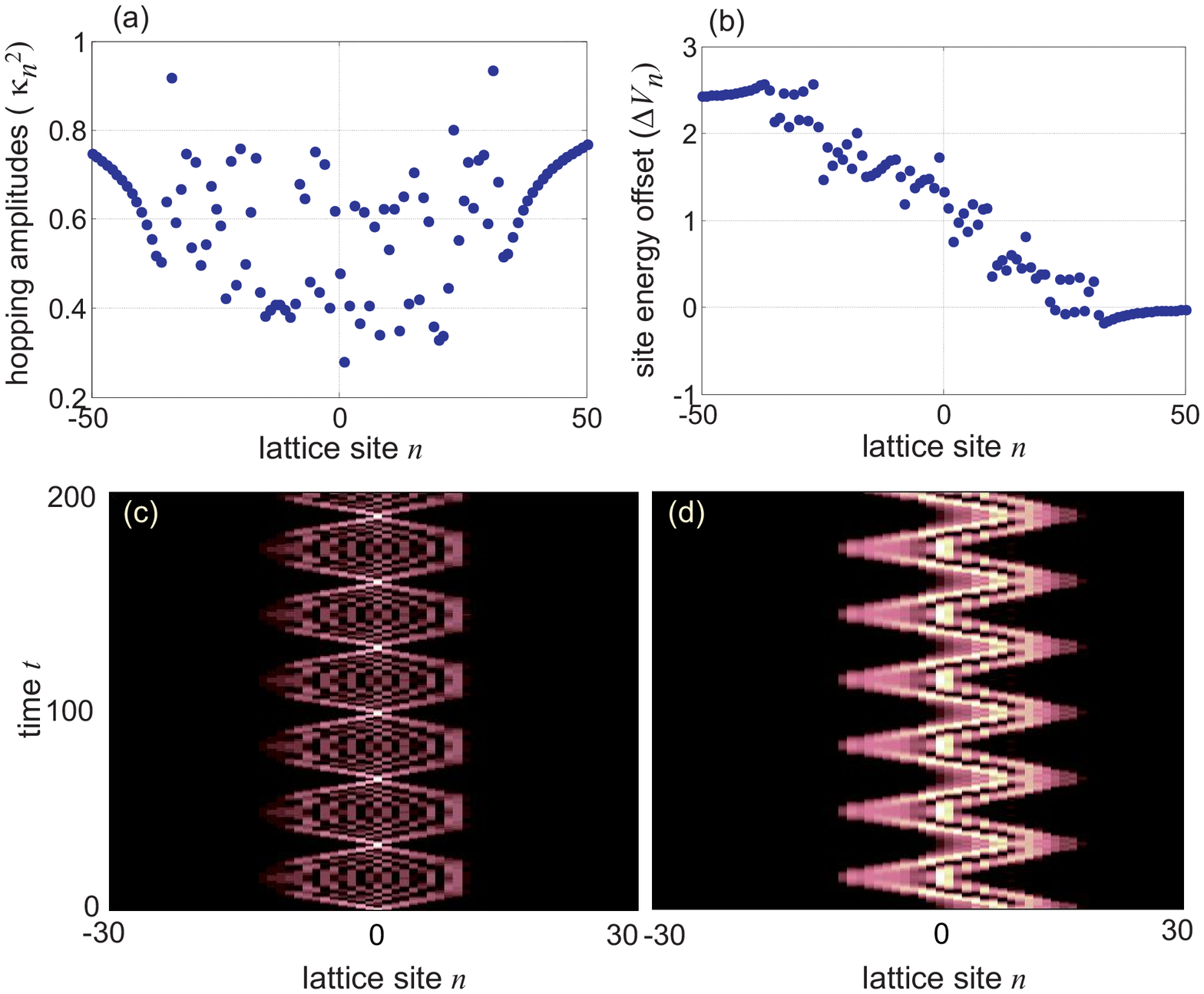}\\
   \caption{(color online) (a) and (b) Same as Fig4., but for the Hamiltonian $\mathcal{H}_{13}$ synthesized from $\mathcal{H}_0$ by
   removing twelve WS resonances (parameter values are $\kappa=1$ and $F=0.2$). In (c) and (d), the persistence of breathing and oscillatory BO modes,
   corresponding to either single-site excitation $c_n(0)=\delta_{n,0}$ [(c)] and to a broad Gaussian wave packet
   excitation $c_n(0)=\exp[-(n-10)^2/64]$ [(d)], are demonstrated.}
\end{figure}
 The persistence of BOs in such synthesized lattices,
corresponding to either single-site excitation at $t=0$ or to a
multiple-site excitation with a broad Gaussian wave packet, is
demonstrated in Figs.3(e-h). In the figures, the
numerically-computed evolutions of $|c_n(t)|^2$ versus $t$, as
obtained by solving the Schr\"{o}dinger equation for the Hamiltonian
$\mathcal{H}_3$ with the initial condition $c_n(0)=\delta_{n,0}$
(for the breathing BO modes, Figs.3(e) and (g)] and
$c_n(0)=\exp[-(n-10)^2/64]$ (for the oscillatory BO modes, Figs.3(f)
and (h)] are depicted for two values of $F$.\\
Tight-binding
lattices with engineered hopping rates and energy sites
corresponding to the ones shown in Fig.3 could be realized using
arrays of evanescently-coupled optical waveguides with engineered
size and distance, in which the distances between adjacent
waveguides control the hopping rates $\kappa_n$ whereas the channel
widths (or refractive index changes) of the guides set the values of
the site energies $E_n$ (see, for instance,
\cite{OBOwaveguides1,Longhi09,Sukhorukov06} and references therein).
\par
The technique of intertwining operators can be iterated by removing
additional resonances from the WS ladder of $\mathcal{H}_0$.
Extended numerical simulations show that the resulting Hamiltonians
turn out to be Hermitian provided that couples of adjacent WS
resonances are removed, whereas in the other cases the hopping rates
can become imaginary at some lattice sites. As a general rule, the
removal of a new WS resonance couple at energies $E=lF, (l \pm 1) F$
introduces in the lattice new defects localized at around the
lattice site $n=l$. For the Hamiltonian $\mathcal{H}_{2s+1}$
obtained by removing from $\mathcal{H}_0$ $s$ couples of adjacent WS
resonances, one has $\kappa_n^{(2s+1)} \rightarrow 1$ for $n
\rightarrow \pm \infty$, $\Delta V_{n}^{(2s+1)} \rightarrow 0$ for
$n \rightarrow + \infty$, and $\Delta V_{n}^{(2s+1)} \rightarrow
2sF$ for $n \rightarrow - \infty$. As an example, Fig.4 shows the
behavior of hopping amplitudes and site energy offsets for a
synthesized lattice obtained by removal of the four WS resonances
with energies $E=0,-F,-25F,-26F$. As the number of removed WS
resonances (and hence of defects in the lattice) increases, the
behaviors of hopping amplitudes $\kappa_n$ and site energies $V_n$
become highly irregular, as shown in the example of Fig.5. Here the
Hamiltonian $\mathcal{H}_{13}$ is synthesized by successive
application of the intertwining operator technique that removes from
the WS ladder spectrum the 12 resonances $E=42F$, $41F$, $26F$,
$25F$, $11F$, $10F$, $0$, $-F$, $-11F$, $-12F$, $-30F$, $-31F$. It
is remarkable that, in such a rather irregular lattice with extended
defects, BOs still
persists and exact self-imaging is attained, as shown in Figs.5(c) and (d).\\
As a final comment, it should be noted that the hopping rates and
site energies of the synthesized lattices described by the Hermitian
Hamiltonians $\mathcal{H}_3$, $\mathcal{H}_5$, $\mathcal{H}_7$, ...
depend on $\gamma$, i.e. on the amplitude of the forcing $F$
entering in $\mathcal{H}_0$. Therefore, a change of the forcing
parameter $F$ in the original WS Hamiltonian $\mathcal{H}_0$ gives
{\em different} lattice realizations for $\mathcal{H}_3$,
$\mathcal{H}_5$, $\mathcal{H}_7$, ...  (see, for instance, Fig.3).
This means that, as an additional term $\mathcal{H}_p=-fn |n
\rangle$ to $\mathcal{H}_0$ simply changes the BO period (just
because the external forcing is changed from $F$ to $F+f$), the
addition of $\mathcal{H}_p=-fn |n \rangle$ to $\mathcal{H}_3$,
$\mathcal{H}_5$, $\mathcal{H}_7$, ... destroys the onset of BOs.
This is shown in Fig.6(a), where the temporal evolution of the site
occupation probabilities $|c_n(t)|^2$ is shown for the Hamiltonian
$\mathcal{H}_{13}$ of Fig.5, with an added perturbation term
$\mathcal{H}_p=-fn|n \rangle$ for $F=0.2$ and $f=-0.1$. For
comparison, the evolution of site occupation probabilities for the
lattice with the WS Hamiltonian $\mathcal{H}_0$ perturbed with
$\mathcal{H}_p=-fn|n \rangle$ (for the same values of $F$ and $f$)
is depicted in Fig.6(b). Therefore, while in a WS ladder Hamiltonian
BOs are observed for any value of forcing $F$ (a change of $F$
corresponds to a change of the BO period), in the synthesized
lattices $\mathcal{H}_3$, $\mathcal{H}_5$, $\mathcal{H}_7$,... a
change of the forcing is detrimental for the onset of BOs. Yet, it
is remarkable that at a fixed forcing strength BOs can be observed
in tight-binding lattices with defects, and even in greatly
irregular lattices (like the one shown in Fig.5).
\begin{figure}[htbp]
  \includegraphics[width=85mm]{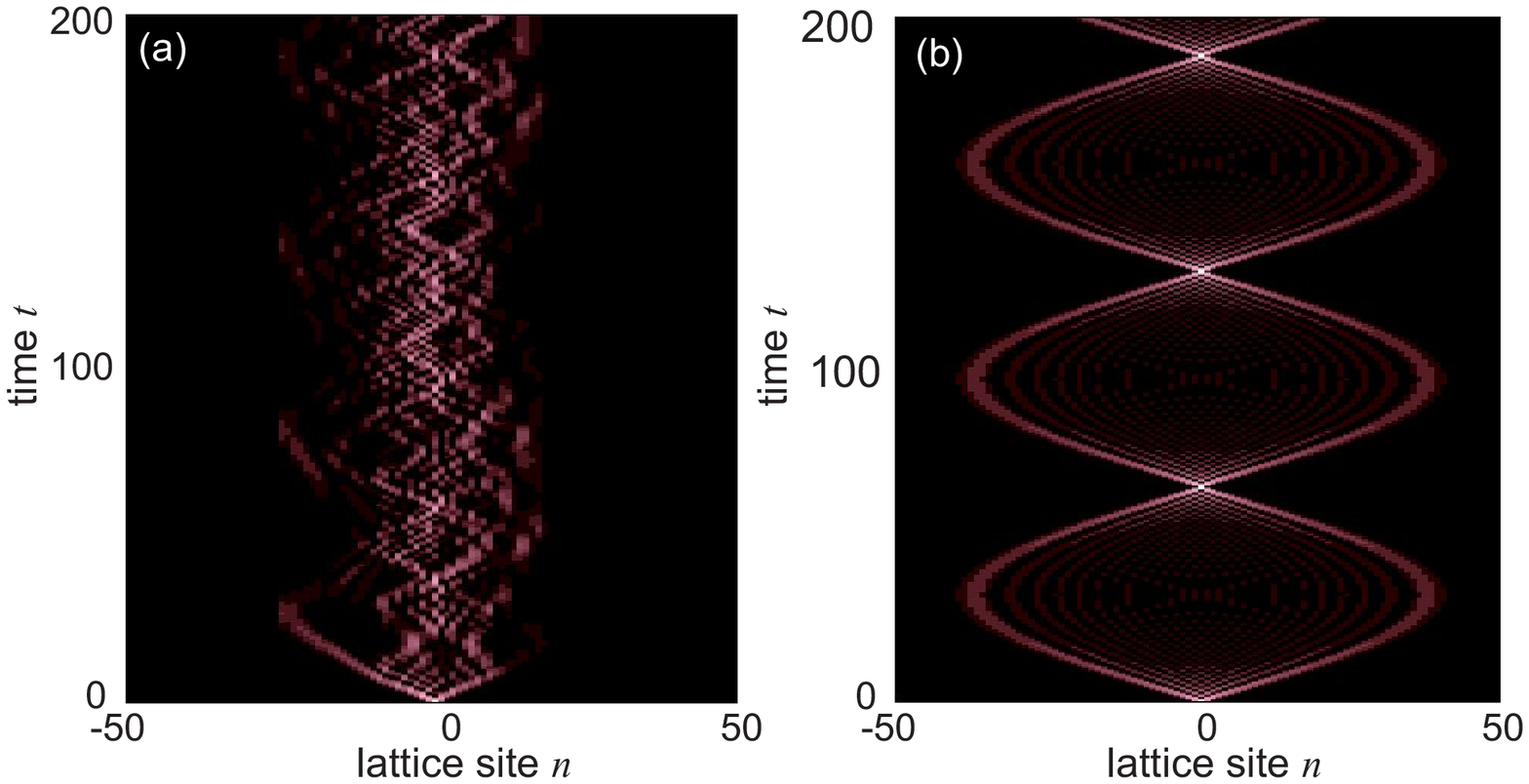}\\
   \caption{(color online) (a) Destruction of BOs (breathing modes) for the lattice Hamiltonian $\mathcal{H}_{13}$ of Fig.5 due to the addition of the
   perturbation $\mathcal{H}_p=-fn |n \rangle$. Parameter values are $\kappa=1$, $F=0.2$ and $f=-0.1$. In (b) the BOs with varied
   period, corresponding to the addition of $\mathcal{H}_p$ to the
   WS lattice Hamiltonian $\mathcal{H}_0$, are shown for comparison.}
\end{figure}

%

\section{Conclusions}
In conclusion, in this work it has been shown that single and
multiple defects in a tight-binding Wannier-Stark lattice can be
introduced such that BOs and the self-imaging property of the WS
lattice are not destroyed. Such lattices are synthesized from the
defect-free lattice by the technique of intertwining operators
generally employed in supersymmetric quantum mechanics to engineer
the spectrum of Hermitian Hamiltonians. The energy spectrum of the
synthesized lattices differs from the Wannier-Stark ladder of the
defect-free lattice because of the missing of pairs of resonances in
the ladder, thus ensuring the persistence of BOs. It is envisaged
that the lattice engineering technique proposed in this work could
be extended to other coherent dynamical regimes, such as dynamic
localization in presence of an ac (time-periodic) force
\cite{Dunlap}. It is also envisaged that the possibility to realize
non-Hermitian tight-binding lattices with real-valued energies
-mentioned in this work- would deserve further investigation and
could stimulate novel studies in the framework of the rapidly
developing field of non-Hermitian quantum mechanics
\cite{note2,Bender}.

\acknowledgments
 This work was supported by the italian MIUR
(PRIN-2008 project "Analogie ottico-quantistiche in strutture
fotoniche a guida d'onda").


\begin{thebibliography}{31}
\bibitem{note1} We assume that the point spectrum of $\mathcal{H}_1$ is
not empty, and that the discrete energy levels are not degenerate,
which is the case of the defect-free Hamiltonian $\mathcal{H}_0$.
\end{thebibliography}

\end{document}